\documentclass[twoside]{article}

\usepackage{PRIMEarxiv}

\usepackage[utf8]{inputenc} 
\usepackage[T1]{fontenc}    
\usepackage{hyperref}       
\usepackage{url}            
\usepackage{booktabs}       
\usepackage{amsfonts}       
\usepackage{nicefrac}       
\usepackage{microtype}      
\usepackage{lipsum}
\usepackage{fancyhdr}       
\usepackage{graphicx}       
\graphicspath{{media/}}     

\title{How to Make Museums More Interactive? \\Case Study of \textit{Artistic Chatbot}
}

\author{\href{https://orcid.org/0009-0005-8473-1402}{\includegraphics[scale=0.06]{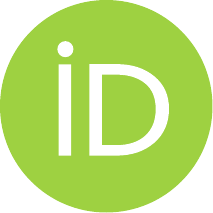}\hspace{1mm}Filip J. Kucia$^{1,2}$},
\href{https://orcid.org/0009-0008-4917-8017}{\includegraphics[scale=0.06]{orcid.pdf}\hspace{1mm}Bartosz Grabek$^1$}, 
\href{https://orcid.org/0009-0004-1402-5580}{\includegraphics[scale=0.06]{orcid.pdf}\hspace{1mm}Szymon D. Trochimiak$^1$}, \href{https://orcid.org/0000-0002-3407-7570}{\includegraphics[scale=0.06]{orcid.pdf}\hspace{1mm}Anna Wróblewska$^1$}
\\
\texttt{filip.kucia@gmail.com}\\
\texttt{\{filip.kucia.stud,bartosz.grabek.stud,szymon.trochimiak.stud,anna.wroblewska1\}@pw.edu.pl}\\
$^1$Faculty of Mathematics and Information Science, Warsaw University of Technology, Warsaw, Poland\\
$^2$Samsung Research and Development Institute Poland, Warsaw, Poland
}

\hypersetup{
pdftitle={How to Make Museums More Interactive? Case Study of Artistic Chatbot},
pdfsubject={},
pdfauthor={Filip J. Kucia, Bartosz Grabek, Szymon D. Trochimiak, Anna~Wr{\' o}blewska},
pdfkeywords={Natural Language Processing, Voice interface, Human-Computer Interaction, Large Language Model},
}

\begin{document}
\maketitle

\begin{abstract}
Conversational agents powered by Large Language Models (LLMs) are increasingly utilized in educational settings, in particular in individual closed digital environments, yet their potential adoption in the physical learning environments like cultural heritage sites, museums, and art galleries remains relatively unexplored. In this study, we present \textit{Artistic Chatbot}\footnote{\url{https://github.com/cinekucia/artistic-chatbot-cikm2025}}, a voice-to-voice RAG-powered chat system to support informal learning and enhance visitor engagement during a live art exhibition celebrating the 15th anniversary of the Faculty of Media Art at the Warsaw Academy of Fine Arts, Poland. The question answering (QA) chatbot responded to free-form spoken questions in Polish using the context retrieved from a curated, domain-specific knowledge base consisting of 226 documents provided by the organizers, including faculty information, art magazines, books, and journals. We describe the key aspects of the system architecture and user interaction design, as well as discuss the practical challenges associated with deploying chatbots at public cultural sites. Our findings, based on interaction analysis, demonstrate that chatbots such as \textit{Artistic Chatbot} effectively maintain responses grounded in exhibition content (60\% of responses directly relevant), even when faced with unpredictable queries outside the target domain, showing their potential for increasing interactivity in public cultural sites. 

During the demo presentation, the audience will be invited to query our \textit{Artistic Chatbot}, which adopts the persona of an artificial art curator, a role that involves responding to questions while simultaneously assessing their relevance to the exhibition.
\noindent The link for the demo video is available \href{https://www.dropbox.com/scl/fi/3ursmsloufobrt04ki1oi/Artistic_Chatbot_Demo_Video.mp4?rlkey=ybq30vrzlmglqgg0673xs3peq&st=ixo6omm7&dl=0}{here}.
\end{abstract}


\section{Introduction}
The rapid advancements in Artificial Intelligence (AI), and specifically Natural Language Processing (NLP), led to the widespread adoption of Large Language Models (LLMs)~\cite{mctear_transforming_2024}. While LLMs can be used in a variety of settings, they are particularly well-suited for the development of conversational agents, also known as chatbots and voice assistants~\cite{gallo_conversational_2024,foosherian_enhancing_2023}. Such systems, powered by state-of-the-art models like GPT, have proven highly effective in education \cite{labadze_role_2023}, where they are fundamental in the development of intelligent tutoring systems and personalized learning platforms. By leveraging their ability to generate meaningful human-like responses, these assistants greatly enhance user interactions and offer much more tailored and context-grounded conversations~\cite{sun_adding_2021}.

Most educational applications of LLM-powered chatbots focus on individualized learning scenarios~\cite{dan_educhat_2023,wang_learnmate_2025}, that is, systems used in isolated, digital environments, and interaction via text-based interfaces. However, relatively little research explores the potential of conversational agents in more social, shared, and interactive settings such as classrooms, or cultural and public education venues, including exhibition sites at museums and galleries~\cite{app11167420,CASILLO2020234}.

Several previous works highlight the effectiveness of using chatbots in cultural heritage and museum sites for enhancing visitor engagement and easier access to information~\cite{10.1145/3341105.3374129,10.1145/3563359.3596661}. A chatbot system deployed in archaeological sites in Pompeii (Italy) used semantic analysis to extract topics from written user queries, contextualize them with information from third-party web services (e.g., Wikipedia, TripAdvisor), and a static knowledge base to formulate answers to visitors' questions~\cite{lombardi2019,SPERLI2021115277}. Another study approached contextual visual question answering with the use of Multi-modal Large Language Models (MLLMs)~\cite{rachabatuni_context-aware_2024}. The system combined visual inputs with the relevant artwork descriptions from a database of Wikipedia content, ensuring scientifically accurate information is provided.

In this work, we demonstrate \textit{Artistic Chatbot}, a voice-to-voice Question-Answering (QA) chatbot developed for and deployed during the exhibition commemorating the 15th anniversary of the Faculty of Media Art at the Warsaw Academy of Fine Arts, Poland. The system allows visitors to ask free-form spoken questions about the faculty, artists, exhibitions, and pieces of art, and receive human-like spoken answers, which are grounded in the domain-specific corpus provided beforehand by the organizers.

\section{System Design and Implementation}
\label{sec:system-design-and-implementation}

The development of the chatbot involved two stages: a data preprocessing phase to construct a specialized knowledge base, and an inference pipeline designed to handle user interactions.

\subsection{Data Preprocessing}

\begin{figure}[!hpt]
    \centering
    \includegraphics[width=0.48\textwidth]{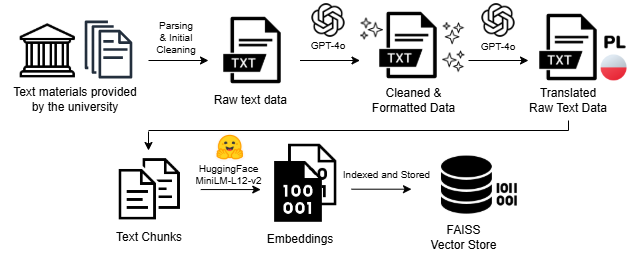}
    \caption{Data Preprocessing Pipeline}
    \label{fig:data-preprocessing-diagram-1}
\end{figure}

The initial stage focused on preparing the knowledge base from a corpus of $226$ ($159$ raw PDF documents and $67$ artists' and academy employees' biographies) provided by the Academy of Fine Arts in Warsaw (see Figure \ref{fig:data-preprocessing-diagram-1}). 
These source documents presented considerable challenges, primarily due to complex layouts, the presence of multiple languages --- such as Polish, Mandarin Chinese, English, German, and French --- and inherent inconsistencies common in unstructured data. While about 92\% of the data was in Polish, many documents included fragments in the other languages, often within the same page, which added to the overall complexity of interpretation and processing.
 
To refine this raw data into a high-quality, consistent resource, the extracted text files were cleaned and translated into Polish using a multilingual Large Language Model -- GPT-4o~\cite{gpt_4o}.

The dataset documents were segmented into $11,596$ smaller, overlapping chunks to facilitate effective Retrieval-Augmented Generation (RAG)~\cite{lewis2021retrievalaugmentedgenerationknowledgeintensivenlp}. The chunks were capped at $5,000$ characters, with an overlap of $200$ characters to ensure contextual continuity between segments. On average, each document was split into approximately $51$ chunks (median: $19.5$), though the number varied depending on length and structure (ranging from $1$ to $650$ chunks per document). Each generated text chunk was then transformed into a dense vector representation (embedding) using the \href{https://huggingface.co/sentence-transformers/paraphrase-multilingual-MiniLM-L12-v2}{\textit{sentence-transformers/paraphrase-multilingual-MiniLM-L12-v2}} model. Subsequently, these embeddings were indexed into a FAISS (Facebook AI Similarity Search)~\cite{douze_faiss_2025} vector store for fast lookup of the most relevant chunks.

\subsection{Inference Pipeline}
\begin{figure*}[!ht]
    \centering
    \includegraphics[width=0.75\textwidth]{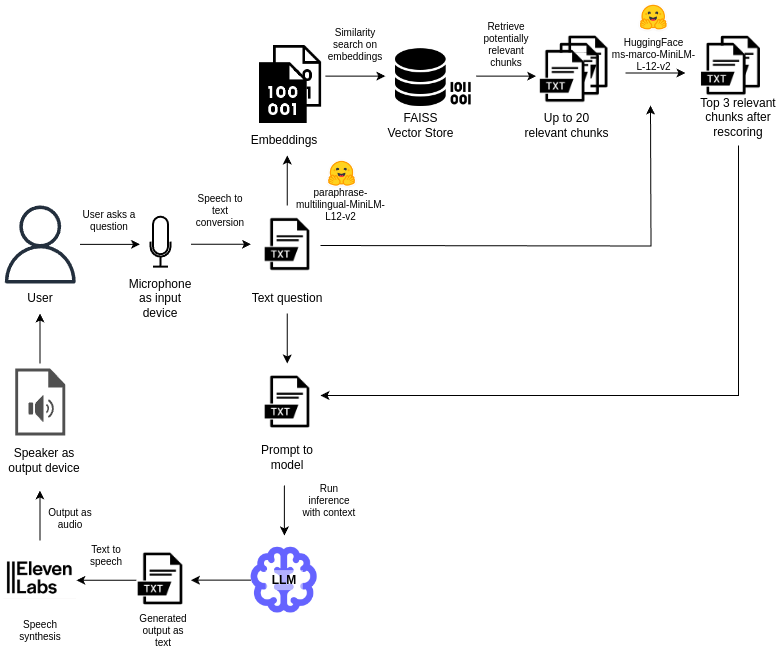}
    \caption{Inference Pipeline}
    \label{fig:data-inference-diagram}
\end{figure*}

The second stage, the inference pipeline, was key to the user interaction flow (see Figure \ref{fig:data-inference-diagram}). User engagement starts with voice input. The system actively listened for predefined trigger phrases (see Section~\ref{sec:interaction}) before capturing the user’s query. This allowed us to avoid undesired triggers and clearly indicated the start of the query. Upon successful query transcription into text, the RAG mechanism was invoked. This involved a two-step retrieval process. First, the user's query was embedded. The FAISS vector store then performed an initial similarity search, retrieving the top $20$ potentially relevant chunks from the indexed knowledge base. This initial retrieval prioritized speed over fine-grained relevance; thus, we needed a subsequent re-ranking step. It employed a pre-trained CrossEncoder model \href{https://huggingface.co/cross-encoder/ms-marco-MiniLM-L12-v2}{\textit{cross-encoder/ms-marco-MiniLM-L12-v2}}, which evaluated the relevance of each of the $20$ retrieved passages against the original query more precisely. Only the top $3$ highest-scoring chunks, as determined by the re-ranker, were selected to form the contextual basis for the final response generation. The selected context, along with the user's query, was then used to construct a prompt for the LLM. The response generation was handled by the GPT-4o-mini model. The prompt included a dynamically selected system message, with a response style chosen randomly at the start of the session, including 'normal', 'academic', or 'laid-back' templates. This system prompt defined the chatbot's persona to differentiate the system's style of answers. The prompt also provided crucial contextual information about the exhibition, such as its location, exhibition period, and outlined specific conversational guidelines, including responding only in Polish and avoiding direct reference to the retrieved context. Finally, the text response was synthesized back into audible speech using the ElevenLabs Text-to-Speech API.
Each complete interaction, including the timestamp, user input query, the style of the system prompt used, and the final generated response, was systematically logged to enable continuous monitoring and subsequent analysis of the chatbot's performance.

\subsection{User Interaction}
\label{sec:interaction}
\begin{figure}[!htbp]
    \centering
    \includegraphics[width=0.49\textwidth]{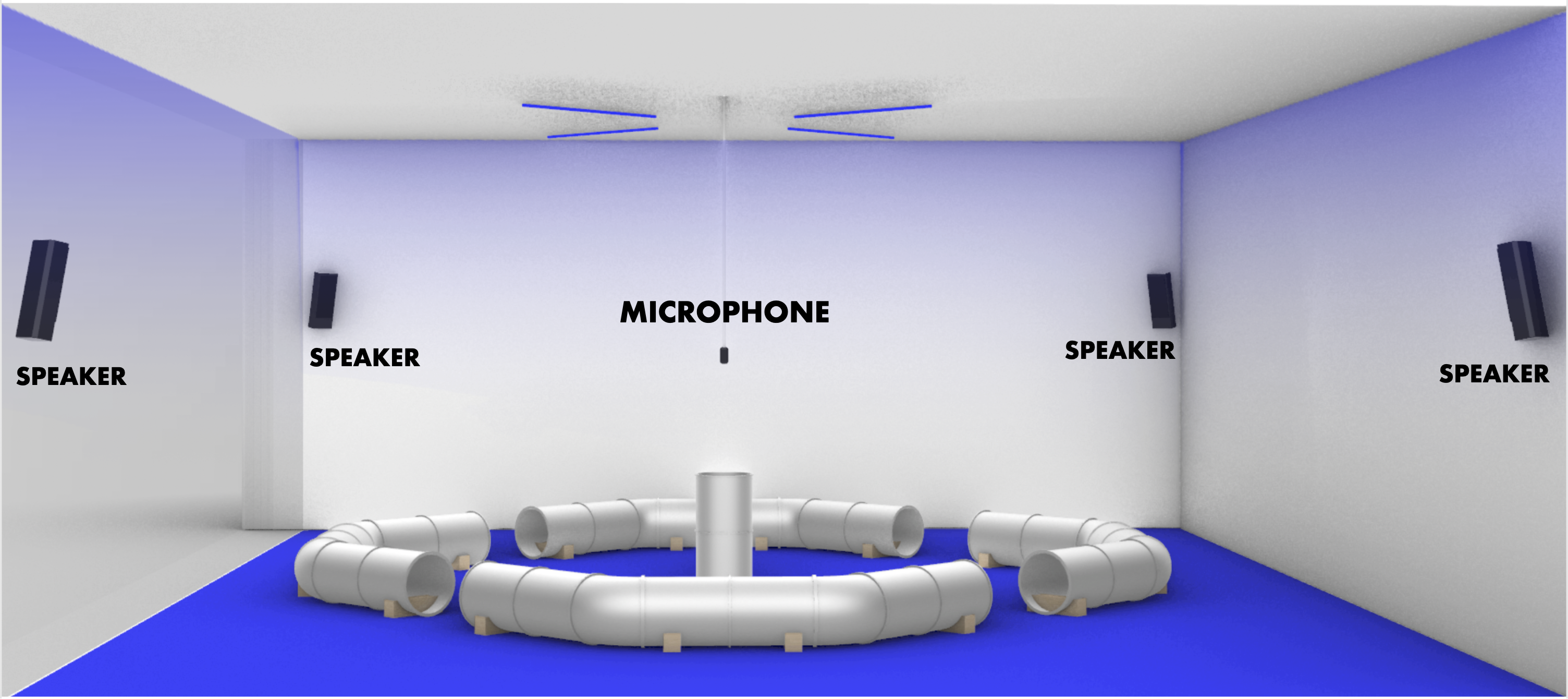}
    \caption{Chatbot Physical Setup}
    \label{fig:exhibition-setup}
\end{figure}

The exhibition physical setup included a ceiling-mounted microphone, located in the center of the room, and four speakers installed in the corners. The peripherals were connected to a nearby PC station running the chatbot and logging user responses in real-time (see Figure~\ref{fig:exhibition-setup}).

To initiate interaction with the \textit{Artistic Chatbot}, visitors were required to walk up to the suspended microphone and articulate one of the four designated \textit{trigger expressions} (``Hello,'' ``Welcome,'' ``Question,'' or ``I have a question''\footnote{Translated from Polish: ``Cześć'', ``Witaj'', ``Pytanie'', ``Mam pytanie''}). Upon recognizing a trigger, the system would greet the visitor, thereby signaling its readiness to receive a question. This mechanism ensured that the input was intentional, distinguishing user interaction from background noise.

The primary aim was to encourage visitors to ask questions related to the exhibition, the participating artists, or the history of the Faculty of Media Art. Once greeted, users could pose virtually any question within that thematic scope. The chatbot was designed for single-turn interactions, meaning it responded to one question at a time without retaining contextual information from previous exchanges. Before accepting a new question, the chatbot had to finish articulating the response, which was generated by the LLM and served in audio form.
Visitors were expected to wait until the spoken response had been completely delivered before initiating another query. This constraint ensured clarity, prevented interruptions, and maintained the coherence of the interaction. 

After each question-and-answer cycle, the system automatically returned to its initial state, listening for trigger phrases to reactivate the chatbot. This interaction model enabled a direct, accessible way for users to engage with the content and concept of the exhibition by facilitating simulated conversations with an artificial art curator through voice input.

\section{Interaction Analysis}
\label{sec:results-and-discussion}
A total of $727$ questions were asked over the course of the month in which the exhibition was open to the public. 
We evaluated the questions and the LLM's responses utilizing LLM-as-a-judge technique (following~\cite{gu2024survey}). We prompted an LLM to check the completeness of the questions, expand them, rate the relevance of the responses, and categorize the questions.
One of the main challenges was question completeness (see Table~\ref{tab:interaction-statistics}). Nearly one-third of all captured user queries were incomplete or prematurely cut off. This issue largely stemmed from the system’s reliance on silence-based end-of-utterance detection. 
However, the analysis showed that correcting the questions usually required only minor edits -- on average about 3.4 characters (within the range 1 -- 6, measured by Levenshtein distance).
\begin{table}[ht]
    \caption{Interaction Statistics: Question (Q.) completeness, Question  relevance and Response (R.) relevance to the exhibition domain.}
    \centering
    \begin{tabular*}{\linewidth}{@{\extracolsep{\fill}}|r|r|r|r|}
    \hline
    \textbf{Metric} & \textbf{Yes (\#)} & \textbf{No (\#)} & \textbf{Yes (\%)} \\
    \hline
    Q. completeness          & $497$ 
    & $230$ 
    & 68.37 \\
    Q. relevance & $142$ & $585$ & $19.53$ \\
    R. relevance   & $440$ & $287$ & $60.52$ \\
    \hline
    \end{tabular*}
    \label{tab:interaction-statistics}
\end{table}
\begin{table}[ht]
    \caption{Question-response relevance score distribution (Mean: $2.66$). Note the scale: 1 (not relevant) to 5 (very relevant)}
    \centering
    \renewcommand{\arraystretch}{1.0}  
    \begin{tabular}{|l|p{0.895cm}|p{0.895cm}|p{0.895cm}|p{0.895cm}|p{0.895cm}|}
    \hline
    \textbf{Score} & \centering $1$ & \centering $2$ & \centering $3$ & \centering $4$ & \centering $5$ \tabularnewline
    \hline
    Count & \centering $286$ & \centering $37$ & \centering $169$ & \centering $110$ & \centering $125$ \tabularnewline
    Percentage (\%) & \centering $39.4$ & \centering $5.1$ & \centering $23.2$ & \centering $15.1$ & \centering $17.2$ \tabularnewline
    \hline
    \end{tabular}
    \label{tab:relevance-score-distribution}
\end{table}

As in other exhibition settings, e.g., in ~\cite{Xu_Howarth_Briggs_Cristianini_2024}, the majority of questions -- about $600$ -- were categorized as simple factual questions, followed by about $150$ casual and confirmation questions, and $24$ hypothetical questions (e.g., ``what would happen if X was true''). This behavioral pattern heavily reduced the number of meaningful interactions with the system and was identified as a key area for improvement. Our visitors frequently strayed from the intended scope of questions, which is a common challenge in public, voice-based installations. In fact, only one-fifth of the questions were fully related to the target domain (see Table \ref{tab:interaction-statistics}). However, as the system consistently supplied the LLM with exhibition-related context retrieved from the knowledge base, many answers remained focused on the exhibition, even when the user's questions were unrelated (see Table \ref{tab:relevance-score-distribution}).

The results point to the system's strength in grounding the responses in the target context, which was of particular value for the exhibition. Nevertheless, we identified a need for further development and measurements in input handling and the relevance of responses, especially in open public settings where high variability of user utterances is expected.

\section{Limitations \& Ethical Considerations}
\label{sec:limitations}

The \textit{Artistic Chatbot} served as a live prototype, and naturally involved some practical shortcuts, resulting in several limitations.

Firstly, the dataset utilized for RAG was relatively small.
Expanding the dataset is one solution, yet it might affect retrieval latency and thus the overall system efficiency. These trade-offs, however, are considered inevitable and typical for RAG model inference \cite{shen2024understandingsystemstradeoffsretrievalaugmented}. That is why designing scalable chatbots in similar settings should focus on effective context utilization through optimal chunking \cite{juvekar2024introducingnewhyperparameterrag} and proper choice of RAG system parameters, for instance, the number of retrieved chunks \cite{vladika2025influencecontextsizemodel}. Additionally, the translation of the source materials into Polish could potentially introduce inaccuracies, mainly due to domain terminology and cultural context shifts. 
As art collections typically span numerous languages, alternative approaches may be used instead of fully translation-based RAG pipelines, for example, Cross-lingual RAG \cite{ranaldi2025multilingualretrievalaugmentedgenerationknowledgeintensive}, where retrieval is performed in the original language and only the most relevant chunks are translated prior to response generation.

Secondly, a few User Experience (UX) issues were observed. The system relies on detecting pauses to determine the end of a user's query, which may lead to premature termination of user input retrieval, particularly in noisy environments or when the user hesitates while speaking.
Changing the interaction from voice-command activation to physical button activation to make the system listen to the user is a simple solution to this problem. An alternative approach, without altering the physical interaction interface, could involve enhancing the existing silence detection with a more sophisticated end-of-utterance (EoU)~\cite{zink_predictive_2024} validation mechanism.

Deployment of LLM-powered chat systems in public places, and especially in museums and cultural heritage sites, requires addressing potential ethical issues.  One of the primary concerns is the truthfulness of LLMs and RAG systems. LLMs can produce hallucinations, often giving plausible yet false or misleading information.
While domain RAG reduces the probability of hallucinations, false information may still occur due to inaccuracies in source documents or LLM's misinterpretation of context \cite{niu2024ragtruthhallucinationcorpusdeveloping}. To prevent hallucination and increase the chatbot's reliability, one may consider fine-tuning the model to acknowledge uncertainty, that is, responding with "I don't know" when needed \cite{chen2024honestaifinetuningsmall}. Another ethical aspect to consider is the safety of chatbot responses. Unless content filtering and moderation techniques are employed, there exists an increased risk of generating harmful or inappropriate content when the system is misused or abused by users, for example, via prompt engineering.

\section{Conclusions}
\label{sec:conclusions}
The deployment of \textit{Artistic Chatbot} at a public art exhibition demonstrated the potential of LLM-powered, voice-based agents to enhance interactivity and visitor engagement in cultural heritage settings. The chatbot was successfully run for a month as part of the exhibition. Over $727$ queries were recorded during the month-long event, with more than $60\%$ of answers judged as related to the exhibition, despite only around $20\%$ of user questions being strictly on-topic. 
This shows the system’s ability to keep responses grounded in exhibition content, even when faced with unpredictable input that is out of the target domain. However, retrieval-based responses also introduced trade-offs, such as limited flexibility and reduced answer diversity when handling vague or ambiguous queries. Challenges like human-chatbot interaction design, simplistic end-of-utterance detection, and moderate overall response relevance (mean score: $2.66$) are identified as key areas for future improvement. 

\section*{Acknowledgments}
We would like to thank our partners at ElevenLabs for providing Text-to-Speech service and the employees of the Academy of Fine Arts for their collaboration.



\begin{thebibliography}{10}

\bibitem{mctear_transforming_2024}
Michael McTear and Marina Ashurkina.
\newblock {\em Transforming {Conversational} {AI}: {Exploring} the {Power} of {Large} {Language} {Models} in {Interactive} {Conversational} {Agents}}.
\newblock Apress, Berkeley, CA, 2024.

\bibitem{gallo_conversational_2024}
Simone Gallo, Fabio Paternò, and Alessio Malizia.
\newblock A conversational agent for creating automations exploiting large language models.
\newblock {\em Personal and Ubiquitous Computing}, 28(6):931--946, December 2024.

\bibitem{foosherian_enhancing_2023}
Mina Foosherian, Hendrik Purwins, Purna Rathnayake, Touhidul Alam, Rui Teimao, and Klaus-Dieter Thoben.
\newblock Enhancing {Pipeline}-{Based} {Conversational} {Agents} with {Large} {Language} {Models}, September 2023.
\newblock arXiv:2309.03748 [cs].

\bibitem{labadze_role_2023}
Lasha Labadze, Maya Grigolia, and Lela Machaidze.
\newblock Role of {AI} chatbots in education: systematic literature review.
\newblock {\em International Journal of Educational Technology in Higher Education}, 20(1):56, October 2023.

\bibitem{sun_adding_2021}
Kai Sun, Seungwhan Moon, Paul Crook, Stephen Roller, Becka Silvert, Bing Liu, Zhiguang Wang, Honglei Liu, Eunjoon Cho, and Claire Cardie.
\newblock Adding {Chit}-{Chat} to {Enhance} {Task}-{Oriented} {Dialogues}, May 2021.
\newblock arXiv:2010.12757 [cs].

\bibitem{dan_educhat_2023}
Yuhao Dan, Zhikai Lei, Yiyang Gu, Yong Li, Jianghao Yin, Jiaju Lin, Linhao Ye, Zhiyan Tie, Yougen Zhou, Yilei Wang, Aimin Zhou, Ze~Zhou, Qin Chen, Jie Zhou, Liang He, and Xipeng Qiu.
\newblock {EduChat}: {A} {Large}-{Scale} {Language} {Model}-based {Chatbot} {System} for {Intelligent} {Education}, August 2023.
\newblock arXiv:2308.02773 [cs].

\bibitem{wang_learnmate_2025}
Xinyu~Jessica Wang, Christine Lee, and Bilge Mutlu.
\newblock {LearnMate}: {Enhancing} {Online} {Education} with {LLM}-{Powered} {Personalized} {Learning} {Plans} and {Support}, March 2025.
\newblock arXiv:2503.13340 [cs].

\bibitem{app11167420}
Yeo-Gyeong Noh and Jin-Hyuk Hong.
\newblock Designing reenacted chatbots to enhance museum experience.
\newblock {\em Applied Sciences}, 11(16), 2021.

\bibitem{CASILLO2020234}
Mario Casillo, Fabio Clarizia, Giuseppe D'Aniello, Massimo {De Santo}, Marco Lombardi, and Domenico Santaniello.
\newblock Chat-bot: A cultural heritage aware teller-bot for supporting touristic experiences.
\newblock {\em Pattern Recognition Letters}, 131:234--243, 2020.

\bibitem{10.1145/3341105.3374129}
Giancarlo Sperl\'{\i}.
\newblock A deep learning based chatbot for cultural heritage.
\newblock In {\em Proceedings of the 35th Annual ACM Symposium on Applied Computing}, SAC '20, page 935–937, New York, NY, USA, 2020. Association for Computing Machinery.

\bibitem{10.1145/3563359.3596661}
Konstantinos Tsitseklis, Georgia Stavropoulou, Anastasios Zafeiropoulos, Athina Thanou, and Symeon Papavassiliou.
\newblock Recbot: Virtual museum navigation through a chatbot assistant and personalized recommendations.
\newblock In {\em Adjunct Proceedings of the 31st ACM Conference on User Modeling, Adaptation and Personalization}, UMAP '23 Adjunct, page 388–396, New York, NY, USA, 2023. Association for Computing Machinery.

\bibitem{lombardi2019}
M.~Lombardi, F.~Pascale, and D.~Santaniello.
\newblock An application for cultural heritage using a chatbot.
\newblock In {\em 2019 2nd International Conference on Computer Applications \& Information Security (ICCAIS)}, pages 1--5, 2019.

\bibitem{SPERLI2021115277}
Giancarlo Sperlí.
\newblock A cultural heritage framework using a deep learning based chatbot for supporting tourist journey.
\newblock {\em Expert Systems with Applications}, 183:115277, 2021.

\bibitem{rachabatuni_context-aware_2024}
Pavan~Kartheek Rachabatuni, Filippo Principi, Paolo Mazzanti, and Marco Bertini.
\newblock Context-aware chatbot using {MLLMs} for {Cultural} {Heritage}.
\newblock In {\em Proceedings of the 15th {ACM} {Multimedia} {Systems} {Conference}}, {MMSys} '24, pages 459--463, New York, NY, USA, 2024. Association for Computing Machinery.

\bibitem{gpt_4o}
{OpenAI}.
\newblock Hello gpt-4o.
\newblock \url{https://openai.com/index/hello-gpt-4o/}, 2025.
\newblock Accessed April 12, 2025.

\bibitem{lewis2021retrievalaugmentedgenerationknowledgeintensivenlp}
Patrick Lewis, Ethan Perez, Aleksandra Piktus, Fabio Petroni, Vladimir Karpukhin, Naman Goyal, Heinrich Küttler, Mike Lewis, Wen tau Yih, Tim Rocktäschel, Sebastian Riedel, and Douwe Kiela.
\newblock Retrieval-augmented generation for knowledge-intensive nlp tasks, 2021.

\bibitem{douze_faiss_2025}
Matthijs Douze, Alexandr Guzhva, Chengqi Deng, Jeff Johnson, Gergely Szilvasy, Pierre-Emmanuel Mazaré, Maria Lomeli, Lucas Hosseini, and Hervé Jégou.
\newblock The {Faiss} library, February 2025.
\newblock arXiv:2401.08281 [cs].

\bibitem{gu2024survey}
Jiawei Gu, Xuhui Jiang, Zhichao Shi, Hexiang Tan, Xuehao Zhai, Chengjin Xu, Wei Li, Yinghan Shen, Shengjie Ma, Honghao Liu, et~al.
\newblock A survey on llm-as-a-judge.
\newblock {\em arXiv preprint arXiv:2411.15594}, 2024.

\bibitem{Xu_Howarth_Briggs_Cristianini_2024}
Zhaozhen Xu, Amelia Howarth, Nicole Briggs, and Nello Cristianini.
\newblock Understanding visitors’ curiosity in a science centre with deep question processing network.
\newblock {\em International Journal of Artificial Intelligence in Education}, 34(3):1072–1101, September 2024.

\bibitem{shen2024understandingsystemstradeoffsretrievalaugmented}
Michael Shen, Muhammad Umar, Kiwan Maeng, G.~Edward Suh, and Udit Gupta.
\newblock Towards understanding systems trade-offs in retrieval-augmented generation model inference, 2024.

\bibitem{juvekar2024introducingnewhyperparameterrag}
Kush Juvekar and Anupam Purwar.
\newblock Introducing a new hyper-parameter for rag: Context window utilization, 2024.

\bibitem{vladika2025influencecontextsizemodel}
Juraj Vladika and Florian Matthes.
\newblock On the influence of context size and model choice in retrieval-augmented generation systems, 2025.

\bibitem{ranaldi2025multilingualretrievalaugmentedgenerationknowledgeintensive}
Leonardo Ranaldi, Barry Haddow, and Alexandra Birch.
\newblock Multilingual retrieval-augmented generation for knowledge-intensive task, 2025.

\bibitem{zink_predictive_2024}
Oswald Zink, Yosuke Higuchi, Carlos Mullov, Alexander Waibel, and Tetsunori Kobayashi.
\newblock Predictive {Speech} {Recognition} and {End}-of-{Utterance} {Detection} {Towards} {Spoken} {Dialog} {Systems}, September 2024.
\newblock arXiv:2409.19990 [eess].

\bibitem{niu2024ragtruthhallucinationcorpusdeveloping}
Cheng Niu, Yuanhao Wu, Juno Zhu, Siliang Xu, Kashun Shum, Randy Zhong, Juntong Song, and Tong Zhang.
\newblock Ragtruth: A hallucination corpus for developing trustworthy retrieval-augmented language models, 2024.

\bibitem{chen2024honestaifinetuningsmall}
Xinxi Chen, Li~Wang, Wei Wu, Qi~Tang, and Yiyao Liu.
\newblock Honest ai: Fine-tuning "small" language models to say "i don't know", and reducing hallucination in rag, 2024.

\end{thebibliography}

\end{document}